**High-pressure growth of fluorine-free $SmFeAsO_{1-x}$ superconducting single crystals**

Hyun-Sook Lee,[1] Jae-Hyun Park,[1] Jae-Yeap Lee,[1] Nak-Heon Sung,[2] Ju-Young Kim,[2] B. K. Cho, [2] Tae-Yeong Koo,[3] Chang-Uk *Jung,[4] and Hu-Jong Lee[1,5*]

[1]*Department of Physics, Pohang University of Science and Technology, Pohang 790-784, Republic of Korea*

[2]*Center for Frontier Materials and Department of Materials Science and Engineering, GIST, Gwangju 500-712, Republic of Korea*

[3]*Pohang Accelerator Laboratory, Pohang 790-784, Republic of Korea*

[4]*Department of Physics, Hankuk University of Foreign Studies, Yongin, Gyeonggi 449-791, Republic of Korea*

[5]*National Center for Nanomaterials Technology, Pohang 790-784, Republic of Korea*

## Abstract

Superconducting single crystals of $SmFeAsO_{1-x}$ *without fluorine doping* were grown at a pressure of 3.3 GPa and a temperature of 1350−1450 ℃ by using the self-flux method. Plate-like single crystals of a few−150 μm in their lateral size were obtained. Single crystals showed the superconducting transition at about 53.5 K with a narrow transition width of 0.5 K. The synchrotron-irradiated x-ray diffractometry patterns and high-angle-annular-dark-field scanning transmission electron microscopy images point to the high quality of the crystals with a well-defined layered tetragonal structure. The chemical compositions of the crystals were estimated by using the electron-probe x-ray microanalysis.

* Corresponding author: hjlee@postech.ac.kr

The recent discovery of a new class of superconducting materials based on FeAs layers, $LaFeAsO_{1-x}F_x$, with a superconducting transition temperature of $T_c$=26 K has attracted much attention.[1] Soon after the discovery, there have been many efforts to increase $T_c$ in $ReFeAsO_{1-x}F_x$ by replacing La with other rare-earth elements (Re=Ce, Sm, Pr, and Nd).[2-5] The highest $T_c$ of 55 K to date was obtained in $SmFeAsO_{1-x}F_x$, which was synthesized under high pressure.[6]

There are many similarities and differences between high-$T_c$ cuprates and $ReFeAsO_{1-x}F_x$, which generates a great deal of scientific interest. Both compounds have a two-dimensional layered structure and the superconducting phase is located close to a magnetic instability phase. It is interesting that $ReFeAsO_{1-x}F_x$ exhibits a relatively high $T_c$ in spite of the presence of the element Fe, which can act as a superconducting pair breaker. Thus, it is highly anticipated that the new FeAs-based superconductors may provide a fresh insight into clarifying the long-debated origin of high-$T_c$ superconductivity.

On the other hand, reports of the physical properties observed in poly-crystalline samples are inconsistent with each other.[7-11] This strongly motivated efforts to grow single-crystals of the material, but to date only three groups have reported the successful growth of single crystals.[13-14] However, even the results from these crystals reveal inconsistencies with each other on some important issues. For instance, both multi-[15-17] and single-band nature[18] was observed. Moreover the observation of a fully gapped order parameter[15,18,19] is contrary to general theoretical expectations.[20-24] Thus, basic consensus has yet to be reached on a number of key issues of the FeAs superconductors, such as pairing symmetry and gap nature.

The large difference in the melting temperatures of the starting compounds (for instance, 1565°C for $Fe_2O_3$, 1530°C for Fe, 1306°C for $SmF_3$, 993°C for FeAs, and 900°C for SmAs) is one of the main obstacles to single-crystal growth. Also the high vapor pressure of As can cause the loss of the element during the growth and raises safety concerns. Thus, we adopted a high pressure synthesis technique that has been routinely used by us for the crystal growth of $MgB_2$, $Sr_{1-x}La_xCuO_2$, and $MgCNi_3$.[25-27]

We developed a series of new steps to establish a reliable crystal-growing scheme. We targeted the compound of *fluorine-free* $ReFeAsO_{1-x}$ for the following reasons. Firstly, the multi-component nature of $ReFeAsO_{1-x}F_x$, which contains five elements, complicates the establishment of the optimal growth conditions of single crystals. In contrast, high-quality single crystals of the double-layered $A_{1-x}K_xFe_2As_2$ (A=Ba, Sr, Ca) superconductors (consisting of only four elements) have been successfully grown by many groups.[28-30] Secondly, the superconducting transition temperature $T_c$ in $ReFeAsO_{1-x}F_x$ single crystals grown under high pressure exhibits a rather wide distribution of values. This $T_c$ depends on the doping concentration, which is determined by both the oxygen and fluorine content.[12-14] Thus, reducing the content of the negative ions to keep the doping concentration stable may lead to a narrowed distribution of $T_c$. Finally, it has been theoretically suggested that producing oxygen vacancies instead of fluorine doping induces more electron carriers in the FeAs layers. This can lead to the lattice shrinkage and the enhancement of the electron density of states, which favor a higher $T_c$.[31,32] In this paper, we report on the growth process of $SmFeAsO_{1-x}$ single crystals, *eliminating fluorine elements*, by using an ultra-high-pressure technique and present an in-depth characterization of the resulting crystals.

For the synthesis of $SmFeAsO_{1-x}$ single crystals, we employed the cubic anvil high-pressure furnace together with the self-flux method. For the first step, SmAs was obtained by reacting Sm chips and As pieces at 600°C for 5 h and then 900°C for 10 h in an evacuated quartz tube. Then, SmAs, $Fe_2O_3$, and Fe were mixed to the nominal composition of $SmFeAsO_{0.85}$ and ground thoroughly. The resulting mixture was pressed into a pellet and inserted into a boron-nitride crucible, which was in turn placed in a cubic pyrophyllite cell equipped with a carbon heater. These processes were carried out in an argon-gas atmosphere for safety against toxic arsenic and to protect the compounds from contamination. The whole assembly was then pressed up to 3.3 GPa at room temperature by using a 14-mm cubic multi-anvil-type press and heated to 1350−1450°C within 30 min. The high-temperature heat treatment lasted for 8−10 h, followed by rapid cooling to room temperature. After the pressure was released, the final bulk was mechanically crushed to segregate the single crystals out of the background flux. Plate-shaped single crystals were obtained ranging from a few microns to up to 150 μm in their lateral size.

Optimum growth conditions were tuned by varying the pressure, the heating temperature, the reaction time, and the cooling rate. In general, slow cooling is favorable to grow sizable single crystals, for instance, of the binary $MgB_2$ superconductor where the presence of impurity phases is not a factor of major concern.[25] However, in the case of $SmFeAsO_{1-x}$, we found that the single phase formed at the reaction temperature tended to separate into several different phases, such as $Fe_xAs_y$ and $Sm_xFe_yAs_z$ during slow cooling. This was confirmed by using energy dispersive X-ray spectroscopy after synthesis. Thus, in our method, the long heat treatment is followed by rapid cooling. As a consequence, most of single crystals turned out to be smaller than 200 μm in their lateral size. Nonetheless, the size of crystals is sufficient for transport measurements[33-36] as well as for key crystal characterizations presented in this study.

The shape and the surface morphology of the crystals were examined using optical microscopy and field-emission scanning electron microscopy (FE-SEM), respectively. The detailed crystal structure was investigated by using synchrotron-irradiated X-ray diffractometry (XRD) and Cs-corrected high-resolution scanning transmission electron microscopy (HR-STEM). The chemical compositions of the crystals were determined using electron-probe X-ray microanalysis (EPMA). The temperature dependence of resistivity of $SmFeAsO_{1-x}$ single crystals was measured using a standard four-probe technique. Special care was taken to select single crystals with flat and clean surfaces to fabricate specimens for the transport measurements. Contact leads were made by using photo and e-beam lithographic patterning on the sample surface. Focused-ion-beam etching and Pt deposition were employed to trim the specimens and to reinforce the connection between the crystal and the contact leads, respectively (see the upper inset of Figure 3 for details of the configuration). Single crystals selected from the same growth batch revealed almost the same superconducting and structural properties.

Optical microscopic images of $SmFeAsO_{1-x}$ single crystals with nominal composition of $x$=0.15 are shown in the inset of Figure 1a. The crystals ranged 5−150 μm in lateral size and 0.5−20 μm in thickness. Most of the crystals were found to have irregular plate-like shapes with flat and shiny surfaces, with smaller crystals displaying cleaner and flatter surfaces. The largest crystal shown in the inset of Figure 1a reveals a rectangular shape and a terraced surface, which presumably reflects the underlying layered tetragonal structure of the crystal. The crystal orientation was determined by using XRD with 6+2 circle diffractometer on $60\times30$ $\mu m^2$-sized rectangular $SmFeAsO_{1-x}$ single crystals. Sharp Bragg peaks of (006) and (114) with a narrow full-width-at-half-maximum of 0.14° and 0.15°, respectively, are seen in Figures 1a and 1b. The solid lines are Gaussian best-fit curves. The results confirm that the direction normal to the rectangular crystal surface is the *c*-axis while the plate-shaped surface is along the *ab*-plane. The XRD-peak analysis leads to the estimation of *a*- and *c*-axis lattice constants as $a$=3.909(1) Å and $c$=8.435(2) Å, respectively.

Figure 2a shows the electron diffraction pattern in a selected area of a $SmFeAsO_{1-x}$ single crystal for the [100] beam direction. The pattern clearly illustrates that the crystal possesses a tetragonal structure, which is consistent with the previous report.[1] Figure 2b shows the high-angle-annular-dark-field scanning transmission electron microscopy (HAADF-STEM) image, which exhibits the real space arrangement of each atomic element constituting the $SmFeAsO_{1-x}$ single crystal. The large and bright dots denote the heaviest element, Sm, and the others indicate As and Fe atoms, as presented in Figure 2b. However, the lightest oxygen atoms were not detected to give information on the oxygen vacancies. The HAADF-STEM image clearly reveals a well-defined structure of alternating SmO and FeAs layers including a wiggle in each layer in the tetragonal unit cell, which directly confirms the structural model of SmFeAsO as schematically illustrated in Figure 2c. The approximate lattice constants, estimated from the HAADF-STEM image, were about $a(=b)$=3.9 Å and $c$=8.4 Å, respectively, which are consistent with the values obtained from the XRD analysis.

The upper inset of Figure 3 shows an FE-SEM image of the specimen prepared for the transport measurement, where a long rectangular single crystal (dimension: 30×3.5×0.8 $\mu m^3$) was patterned into a four-probe configuration. The main panel presents the in-plane resistivity of the $SmFeAsO_{1-x}$ single crystal. As shown in the lower inset of Figure 3, the onset of the superconducting transition of the single crystal occurs at about 53.5 K with a very narrow transition width of $\Delta T_c$=0.5 K (by adopting the criterion of 10−90% of the normal-state resistivity). Our value of $T_c$ is consistent with that reported for a polycrystalline sample with the nominal composition of $SmFeAsO_{0.85}$.[37] The transition width $\Delta T_c$ of our $SmFeAsO_{1-x}$ single crystal is significantly smaller than the 1−4 K width reported for fluorine-doped $NdFeAsO_{0.82}F_{0.18}$ single crystals[13,16] determined by the resistive transition. The residual resistivity ($\rho_0$) at $T_c$ of our single crystal is about 0.08 mΩcm, which is also 3-4 times smaller than the $\rho_0(T_c)$=0.35 mΩcm for polycrystalline $SmFeAsO_{0.85}$[37] and $\rho_0(T_c)$=0.28 mΩcm for single crystalline $NdFeAsO_{0.82}F_{0.18}$.[13,16] The corresponding residual resistivity ratio, $RRR \equiv \rho(300\ K)/\rho(T_c(\text{onset})) \approx 5$, of our crystal turns out to be twice as large as the value of ~2.5 seen previously for $NdFeAsO_{0.82}F_{0.18}$ single crystals.[13,16] Since transport properties such as *RRR* sensitively depend on the quality of a specimen, our $SmFeAsO_{1-x}$ single crystal is considered to be of high quality.

The resulting doping level of the single crystal may differ from the designed nominal value. The actual doping of our $SmFeAsO_{1-x}$ single crystals with nominal composition of $x$=0.15 was examined by using the EPMA study, which is one of the best methods to determine the composition ratio of micron-sized samples like our single crystals. Several crystals obtained from the same batch were examined. For higher accuracy we adopted the standard materials Sm (100%), Fe (100%), GdAs (44.2%, 51.8%), and $Al_2O_3$ single crystals (Micro-Analysis Consultants Ltd.). The measurements yielded the composition ratio of Sm:Fe:As:O=1:1:1:0.6 ± 0.1, corresponding to ~40% oxygen vacancy.

We cross-checked the actual doping level of our single crystals by considering the change of the lattice constant for varying oxygen vacancies. According to the report for *polycrystalline* $NdFeAsO_{1-x}$[38] synthesized at high pressure, the *a*- and *c*-axis lattice constants decrease with increasing the oxygen vacancies in $NdFeAsO_{1-x}$ and saturate at the reduction rates of 0.63 % and 0.55 %, respectively, for $x$=0.4 compared to $x$=0. With further increases of the nominal doping level in the range $x$=0.4−0.8, the values of $T_c$ and the *a*- and *c*-axis lattice constants remain unchanged. The corresponding superconducting transition temperature ($T_c$~53 K) of this compound is highest in this doping range. Moreover, the superconducting volume fraction increases with increasing $x$ and reaches a maximum at $x$=0.4 while it begins to decrease for further doping. From these results it was suggested that[38] the composition ratio of Nb:Fe:As:O=1:1:1:0.6 corresponds to the most stable phase in this compound without fluorine doping. Compared to the values of the lattice constants $a$=3.933(5) Å and $c$=8.495(3) Å for undoped SmFeAsO,[6,37] the reduction rates of the lattice constants for our single crystals [$a$=3.909(1) Å and $c$=8.435(2) Å] are 0.61% and 0.71%, respectively. The amount of reduction in *a*- and *c*-axis lattice constants due to 40% oxygen vacancy in our *$SmFeAsO_{1-x}$ single crystal* was quite comparable to those observed in the *polycrystalline $NdFeAsO_{1-x}$* with the same amount of oxygen vacancy. The slightly higher reduction rate for the *c*-axis lattice constant in $SmFeAsO_{1-x}$ than in $NdFeAsO_{1-x}$ is consistent with the earlier report[1] that, in $ReFeAsO_{1-x}$, the lattice shrinks more along the *c*-axis than along the *a*-axis upon putting a smaller-size element to the Re site [the atomic radius of Sm (1.13 Å) is smaller than that of Nd (1.15 Å)]. In addition, the composition ratio of Sm:Fe:As:O=1:1:1:0.6±0.1, estimated from the EPMA measurements, turns out to be almost the same as the one providing the highest superconducting volume fraction in $NdFeAsO_{1-x}$ compound in its polycrystalline state. From these considerations, we conclude that the $x$=0.4 phase is most stable in the fluorine-free compound of $ReFeAsO_{1-x}$ and that single crystals grown under high pressure are most likely to be stabilized in the composition of Sm:Fe:As:O=1:1:1:0.6, as was found for the single crystals in this study.

In summary, we present the successful growth of nominal composition $SmFeAsO_{0.85}$ single crystals *without fluorine elements* under high pressure. The problems arising from the large difference in the melting temperatures of the constituent elements, leading to a high possibility of losing As elements during the synthesis, were overcome in the extreme-high-pressure growth process in a GPa range. The rapid cooling after the long heat treatment prevents the phase separation that often plagues the slow cooling process and, thus, helps maintain the single phase. The high-quality XRD peaks and HR-TEM images indicate the good crystallinity of our single crystals, which have a well-defined layered tetragonal structure. The $T_c$ of our single crystals was 53.5 K with a very narrow $\Delta T_c$ of 0.5 K. In addition, low residual resistivity ($\rho_0$) and especially large residual resistivity ratio (*RRR*) values compared with the reported values for $ReFeAsO_{1-x}F_y$ (Re=Sm, Nd) single crystals indicate the high-quality our single crystals. Growing FeAs-based single crystals without involving fluorine elements provides an easier route to finding the optimum conditions for crystal formation and a simpler interpretation of the electronic structure. This may also simplify finding the gap structure and the pairing symmetry of the superconductivity in these materials, helping to reveal the mechanisms of superconductivity in this family of materials.

*Acknowledgments*

We are grateful to L. Paulius for the critical reading of the manuscript. This work was supported (for HJL) by the Korea Science and Engineering Foundation (KOSEF) through Acceleration Research Grant R17-2008-007-01001-0 and Pure Basic Research Grant R01-2006-000-11248-0, and by Korea Research Foundation (KRF) through Grant KRF-2005-070-C00055. This work was also supported by the Steel Science Program of POSCO, Korea. B. K. Cho was supported by the NCRC program administered by the KOSEF (Grant No. R15-2008- 006-01002-0). C.-U. Jung was supported by KRF (Grant No. KRF-2007-331-C00100) and by KOSEF (Grant No. R01-2006-000-11071-0).

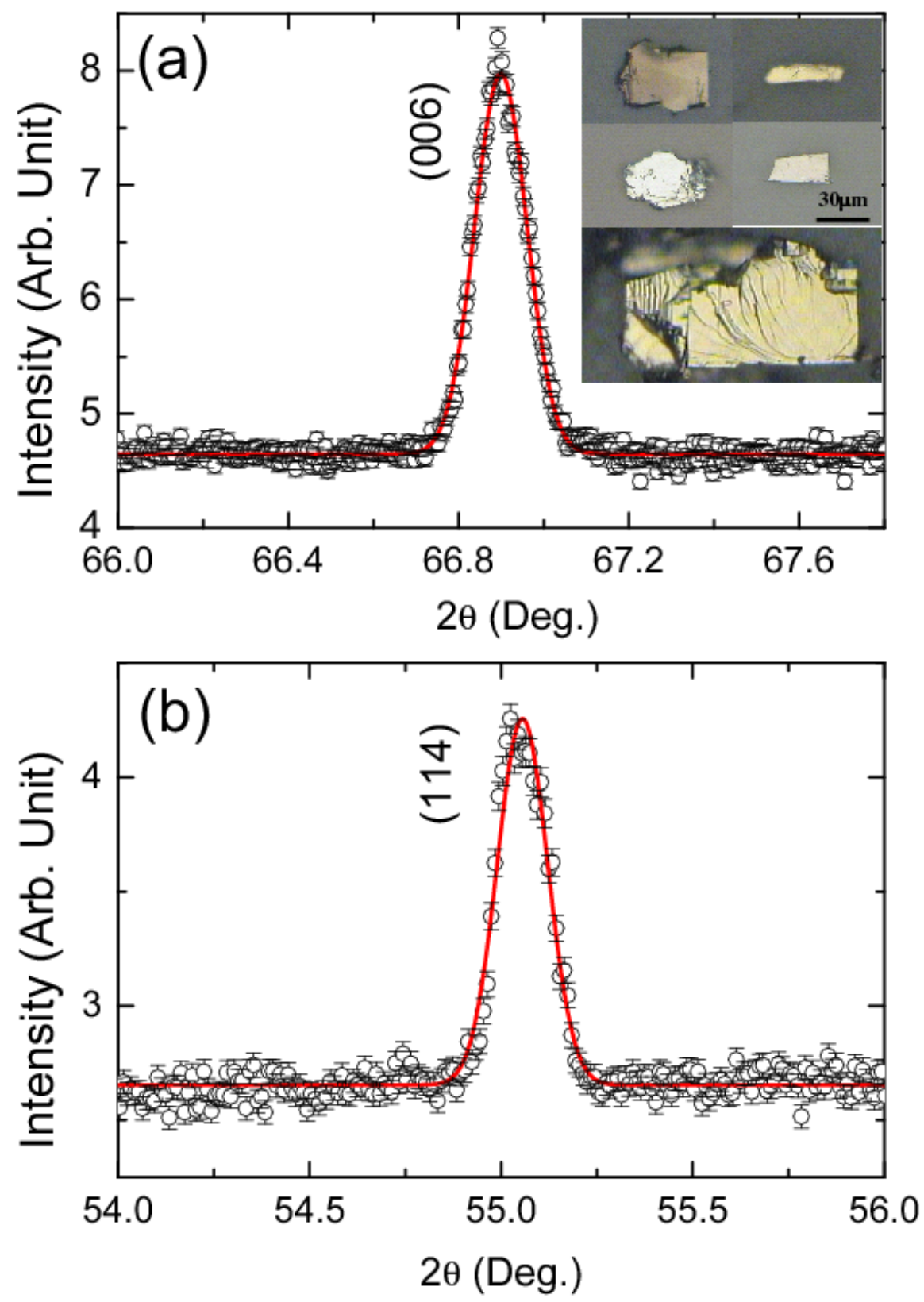

**Figure 1.** Synchrotron-irradiated X-ray diffractometry (XRD) spectra of single crystalline $SmFeAsO_{1-x}$ with nominal composition of $x$=0.15. The theta-2theta scan of (a) (006) and (b) (114) peaks display excellent crystallinity. Solid lines are Gaussian best-fit curves. Inset: optical microscopy image of $SmFeAsO_{1-x}$ single crystals grown from the self-flux method under high pressure.

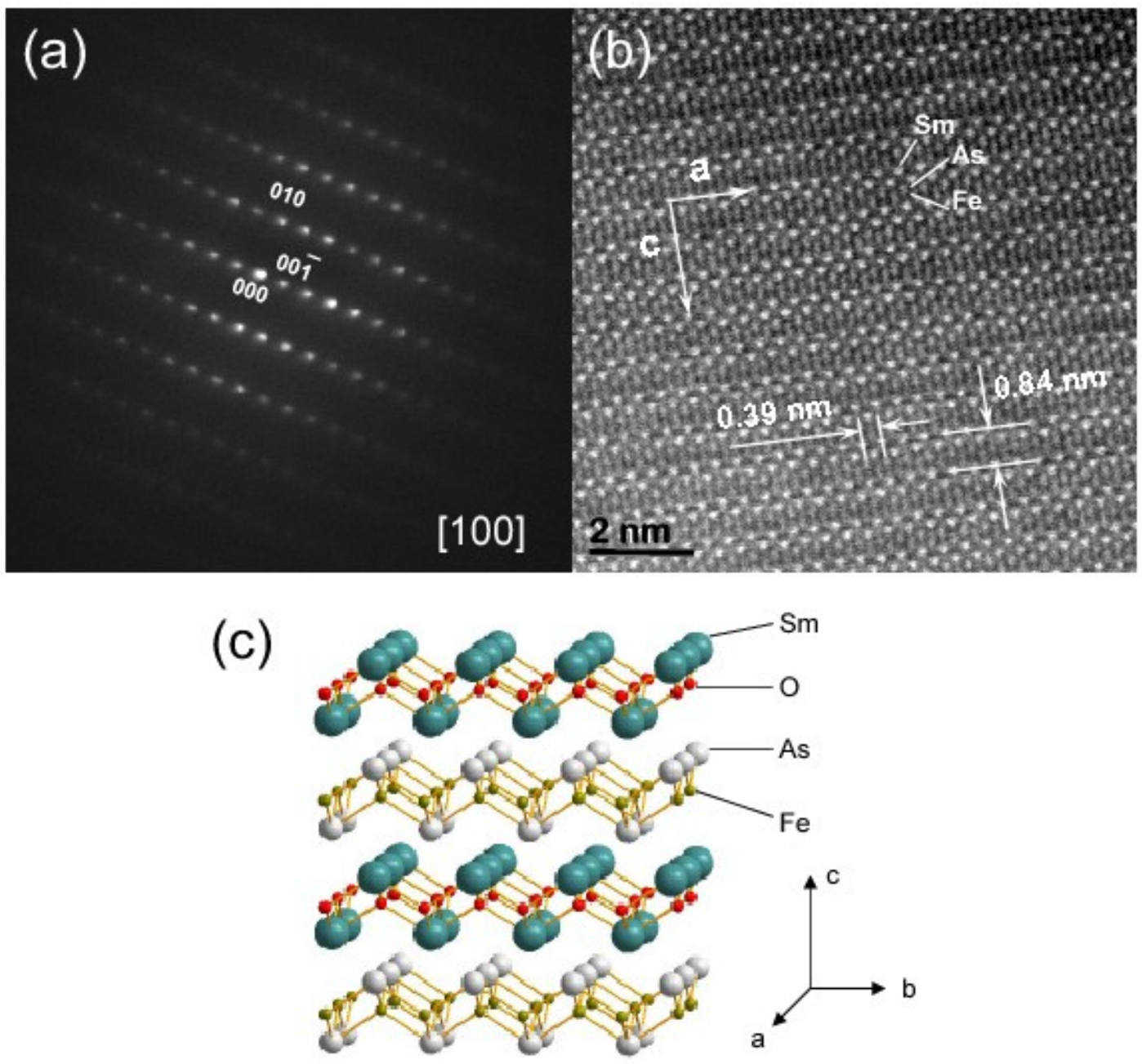

**Figure 2.** Transmission electron microscopy images of a $SmFeAsO_{1-x}$ single crystal obtained by using Cs-corrected high-resolution scanning transmission electron microscopy (HR-STEM). (a) A selected-area electron diffraction pattern for the beam direction of [100] and (b) a high-angle-annular-dark-field scanning transmission electron microscopy (HAADF-STEM) image, which show a well-defined structure of alternating SmO and FeAs layers. Focused-ion-beam etching was employed to prepare the crystal for the imaging above. The images in (a) and (b) are consistent with the tetragonal ZrCuSiAs-type structure of the SmFeAsO compound as schematically illustrated in (c).

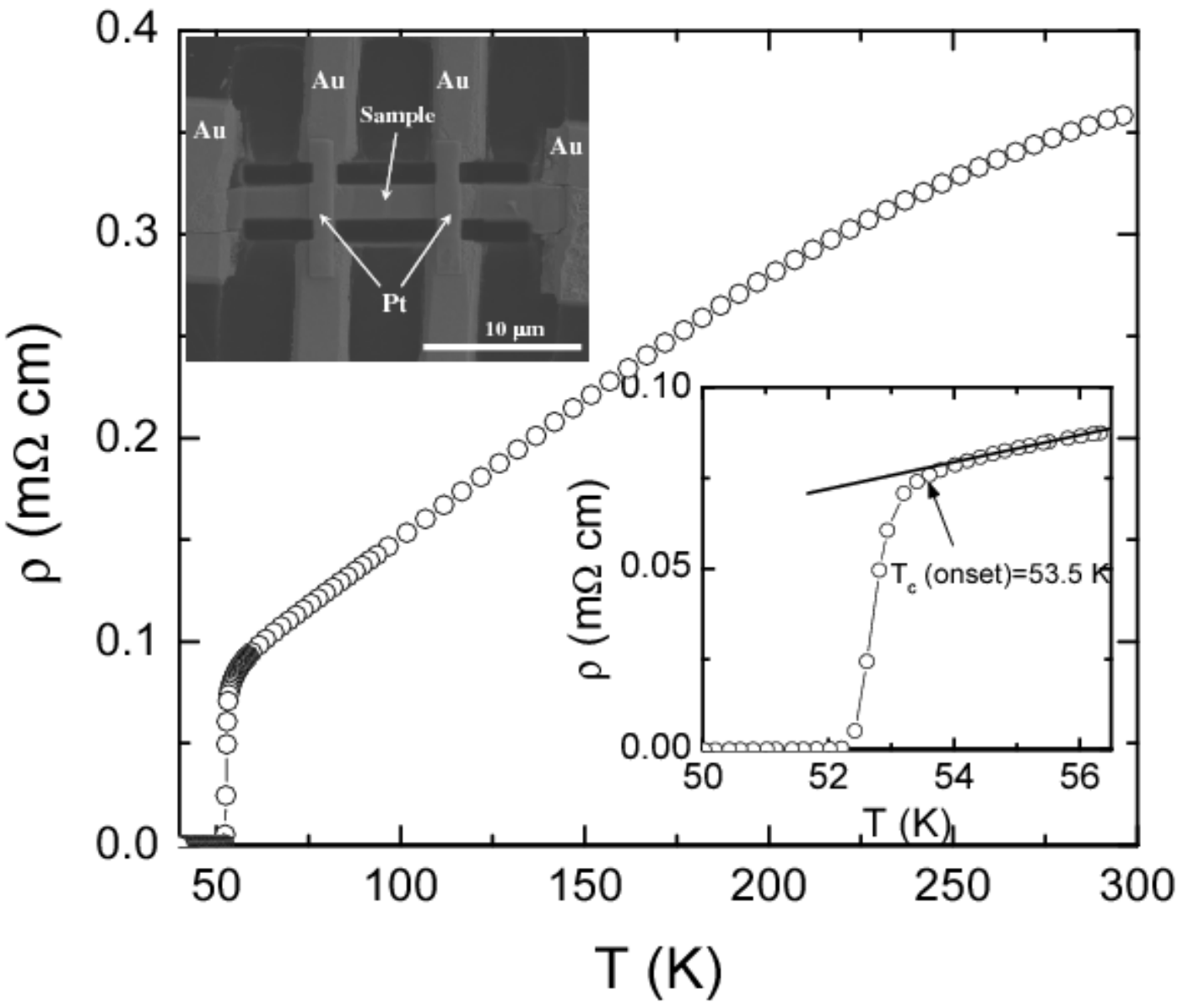

**Figure 3.** Temperature dependence of the resistivity in zero magnetic field for the $SmFeAsO_{1-x}$ single crystal. The upper inset shows the field-emission scanning electron microscopy (FE-SEM) image of the four-probe patterned crystal for measurements of the resistive transition. A rectangular single crystal of dimension 30×3.5×0.8 $\mu m^3$ was patterned into a configuration with the spacing between the voltage probes and the width of the specimen after trimming the crystal side edges by a focused ion beam were 6.3 μm and 2.4 μm, respectively. The inset shows a close-up view of the data near the superconducting transition temperature, denoting the onset of deviation [$T_c$(onset)] from the linearity.